\documentclass[10pt,aps,amsmath,amssymb,prl,twocolumn,superscriptaddress,longbibliography]{revtex4-1}

\usepackage[colorlinks=true,linkcolor=blue,citecolor=blue]{hyperref}
\usepackage{amsfonts}
\usepackage{dcolumn}
\usepackage{bm}
\usepackage{tikz}
\usepackage{amsmath,amsfonts,amssymb,times,natbib}
\usepackage{multirow, makecell}
\usepackage[standard]{ntheorem}
\usepackage{graphicx}
\usepackage{graphics}
\usepackage{amssymb}
\usepackage{siunitx}
\usepackage{amsmath}
\usepackage{amssymb}
\usepackage{setspace}
\usepackage{float}
\usepackage{natbib}
\usepackage{lineno}
\usepackage{setspace}

\usepackage{changes}

\begin{document}

\title{Learning Entanglement Quasiprobability from Noisy and Incomplete Data}
\author{Yu-Zhuo Li}
\affiliation{MIIT Key Laboratory of Complex-field Intelligent Sensing, School of Optics and Photonics, Beijing Institute of Technology, Beijing 100081, China}
\author{Li-Chao Peng}
\email[Contact author: ]{plc@bit.edu.cn}
\affiliation{MIIT Key Laboratory of Complex-field Intelligent Sensing, School of Optics and Photonics, Beijing Institute of Technology, Beijing 100081, China}
\affiliation{Yangtze Delta Region Academy of Beijing Institute of Technology (Jiaxing), Jiaxing 314019, China}
\affiliation{Center for Photonic Quantum Precision Measurement, Advanced Research Institute of Multidisciplinary Science, Beijing Institute of Technology, Beijing 100081, China}
\author{Ke-Mi Xu}
\email[Contact author: ]{xukemi@bit.edu.cn}
\affiliation{MIIT Key Laboratory of Complex-field Intelligent Sensing, School of Optics and Photonics, Beijing Institute of Technology, Beijing 100081, China}
\affiliation{Yangtze Delta Region Academy of Beijing Institute of Technology (Jiaxing), Jiaxing 314019, China}
\affiliation{Center for Photonic Quantum Precision Measurement, Advanced Research Institute of Multidisciplinary Science, Beijing Institute of Technology, Beijing 100081, China}

\begin{abstract}
Negativities in quasiprobability distributions, a foundational concept originating in quantum optics, serve as a fundamental signature of quantum nonclassicality, with entanglement quasiprobabilities offering a necessary and sufficient criterion for entanglement. However, practical reconstruction of entanglement quasiprobabilities conventionally requires full quantum state tomography, severely limiting scalability. Here, we propose a deep-learning framework that reconstructs entanglement quasiprobabilities directly from incomplete local projective measurements, bypassing full state reconstruction. Using a residual neural network, partial measurement outcomes are mapped to high-fidelity entanglement quasiprobabilities. Numerical benchmarks up to three qubits show more than a $30\times$ reduction in reconstruction error compared with state-of-the-art tomographic methods. Experimental validation on photonic entangled states demonstrates reconstruction and entanglement detection with substantially reduced measurement resources. Our results establish machine-learning-assisted reconstruction of entanglement quasiprobabilities as a scalable and practical tool for entanglement characterization in quantum optical systems.
\end{abstract}

\maketitle
\section{\label{sec:introduction}Introduction}
Quantum entanglement lies at the heart of quantum mechanics~\cite{EPR1935,Schrödinger1935}. As a central resource in quantum information science, it enables transformative technologies spanning quantum computation~\cite{raussendorf2001one}, communication~\cite{gisin2007quantum}, and metrology~\cite{giovannetti2006quantum}. Harnessing these capabilities in practice critically relies on reliable certification of entanglement, which remains a key challenge for pushing quantum technologies beyond classical boundaries~\cite{Horodecki2009Quantum}.

The measurement and characterization of quantum correlations are also essential for understanding many-body quantum systems~\cite{Kunkel2022Detecting, Brunner2022Many, schweigler2017experimental}, optimizing quantum computing protocols~\cite{Talsma2024Continuously, Santra2025Genuine}, and verifying quantum advantage~\cite{KahanamokuMeyer2022Classically}. Over the years, a broad range of approaches have been developed to detect and quantify entanglement, including the Peres–Horodecki criterion~\cite{Peres1996Separability,PawelI1997Separability}, entanglement witnesses~\cite{Barbara2000Bell}, concurrence~\cite{Horodecki2009Quantum}, mutual information~\cite{Valdez2017Quantifying}, and entanglement entropy based on the von Neumann entropy~\cite{Bennett1996Mixed}. However, these methods often face practical limitations, such as incomplete access to state information, sensitivity to experimental noise, and poor scalability to high-dimensional or many-body systems~\cite{Horodecki2009Quantum}. Nevertheless, these advances have enabled landmark experimental milestones, ranging from violations of two-particle Bell inequalities~\cite{Aspect1982Experimental} to the generation of multipartite entangled states~\cite{Pan2001Experimental}, and have demonstrated growing potential for applications in quantum networks~\cite{pan2012multiphoton}.

Entanglement quasiprobability (EQP)~\cite{Sanpera1998Local, Vidal199Robustness, Sperling2009Representation} provides a framework for characterizing quantum entanglement across diverse physical platforms. In general, quasiprobability distributions originated in quantum optics as a fundamental tool for characterizing nonclassical effects. The most prominent example is the Glauber–Sudarshan P function \cite{glauber1963coherent,sudarshan1963equivalence}, whose negativities certify that a non-classical state cannot be described as a classical mixture of coherent states. A key feature of EQPs is that negativities in the distribution constitute a necessary and sufficient criterion for entanglement certification. Moreover, the complete set of EQPs can be used to reconstruct the underlying density matrix, thereby enabling full quantum-state characterization. Recent experimental progress has enabled the direct reconstruction of EQPs for photonic Bell states~\cite{Sperling2019Experimental}. Optimization-based methods have further confirmed the completeness of the EQP description within the full Hilbert space~\cite{Prasannan2021Experimental}, non-Gaussian continuous-variable states and multimode systems~\cite{Sperling2012Entanglement, Bohmann2017Entanglement, Sperling2018Quasiprobability}. Nevertheless, practical EQP reconstruction remains severely constrained by the prohibitive measurement overhead of full quantum state tomography and its associated computational complexity, limiting its applicability to larger-scale quantum systems.
\begin{figure*}[t]
\centering
\includegraphics[scale=0.23]{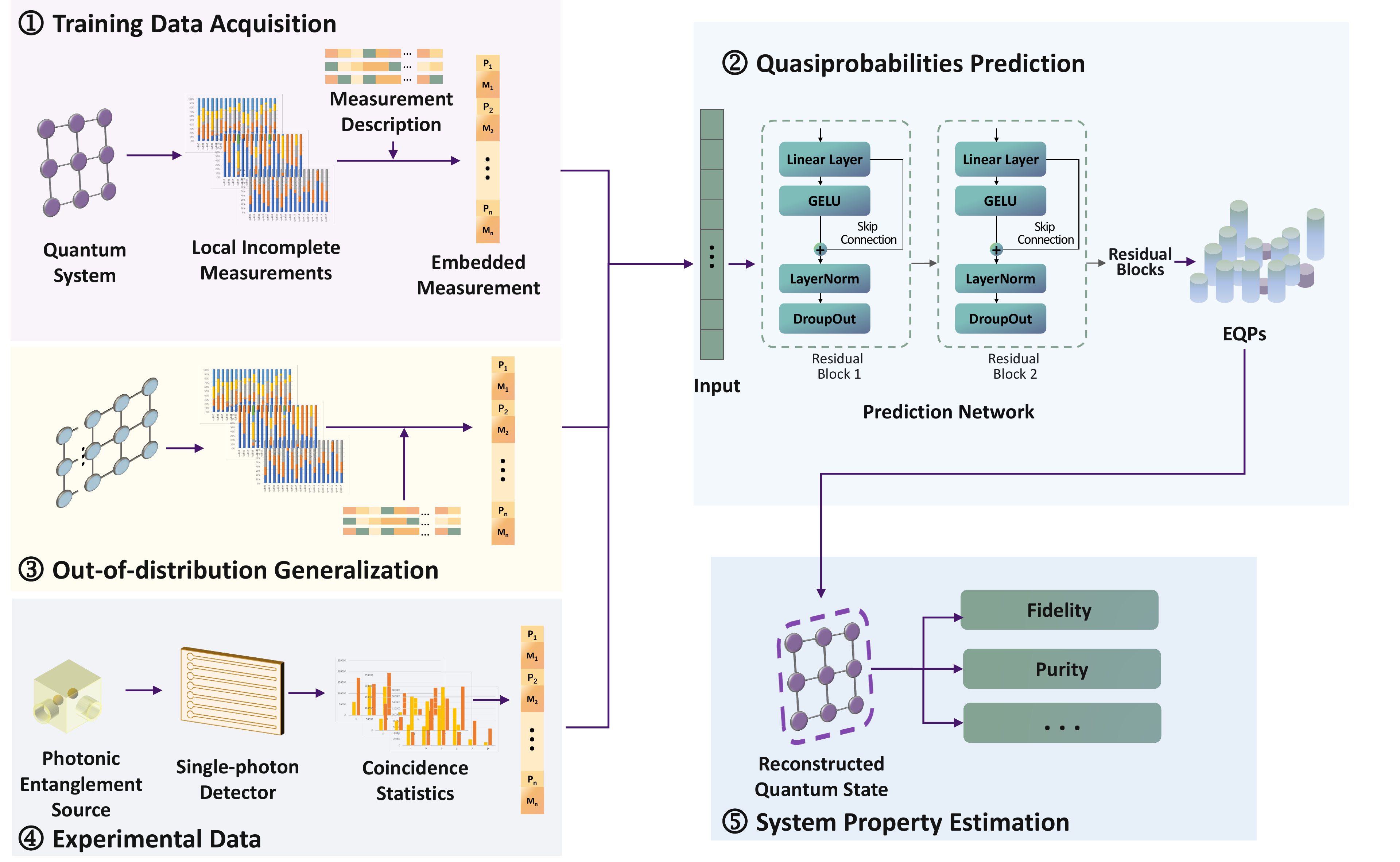}
\caption{\textbf{Architecture of deep-learning-assisted EQP reconstruction from incomplete measurements}. Incomplete local measurement outcomes (1) are mapped by a trained network to entanglement quasiprobabilities (2). The framework exhibits strong generalization to out-of-distribution states (3), robustness to experimental noise (4), and enables accurate estimation of fidelity, purity, and other state properties (5).}
\label{fig:framework}
\end{figure*}

Numerous techniques have been developed to reduce the number of required measurement settings, including randomized benchmarking~\cite{Knill2008Randomized}, compressed-sensing-based quantum state tomography~\cite{Gross2010Quantum}, direct fidelity estimation~\cite{Flammia2011Direct}, self-testing approaches~\cite{Bancal2015Physical}, adaptive quantum state verification~\cite{Takeuchi2018Verification}, optimal entanglement verification protocols~\cite{Pappa2012Multipartite, McCutcheon2016Experimental, Pallister2018Optimal, Barreiro2013Demonstration}, and shadow tomography~\cite{aaronson2018shadow, huang2020predicting}. Most of these methods focus on estimating specific parameters, such as the expectation value of an entanglement witness, which are subsequently compared against predefined thresholds to determine whether a given quantum state is entangled. Despite these advances, the minimal amount of information required to faithfully represent a generic quantum state with a prescribed accuracy remains an open question~\cite{Rocchetto2019Experimental, Gebhart2023Learning}.

To address this challenge, neural-network-based methods have emerged as a powerful approach for estimating quantum properties from sparse measurement data~\cite{ma2025machine, Gaikwad2024Neural, Dominik2023Deep, aguilar2015experimental}. In particular, schemes based on incomplete measurements provide a scalable alternative to full quantum state tomography. By targeting selected observables, these methods can dramatically reduce measurement overhead~\cite{aaronson2018shadow, Paris2004Quantum} while still enabling the inference of local or global properties of quantum systems from a substantially reduced dataset. Such approaches are especially relevant for noisy intermediate-scale quantum (NISQ) devices, where experimental resources are inherently limited and efficient characterization protocols are essential~\cite{Torlai2018Neural, Carrasquilla2017Machine}. Nevertheless, most existing methods are primarily designed for entanglement verification and provide limited access to other physical properties of the quantum state. A general framework capable of directly reconstructing the full density matrix—thereby enabling the simultaneous prediction of entanglement and other key quantum characteristics—has so far remained an unresolved challenge.

Here, we introduce the Entanglement Quasiprobability Prediction Network with Adaptive Incomplete Measurement Encoding (EQPs-AIME-Net) as shown in Fig.~\ref{fig:framework}, a deep-learning framework that directly reconstructs entanglement quasiprobabilities from incomplete and noisy local measurements. Our approach aims to mitigate the experimental overhead associated with full quantum state tomography and to investigate whether entanglement quasiprobabilities can be inferred directly from incomplete local measurements. This study further demonstrates that the inferred EQPs can serve as a compact and physically transparent representation, potentially enabling both the certification of nonclassical correlations and the prediction of multiple state properties within a unified framework.
\section{\label{sec:results}Results}
We present numerical and experimental results for reconstructing EQPs from incomplete measurements. Performance is evaluated for two- and three-qubit systems and compared with likelihood-based baselines. We consider scenarios where EQP reconstruction is performed without complete measurement bases and where computational resources are constrained. The proposed deep neural network approach is applied to investigate the feasibility of reducing measurement requirements while maintaining accurate EQP quantification. We further explore its applicability to complex systems, focusing on the reconstruction of multi-qubit quantum states with access to both entanglement characteristics and global properties under limited measurement settings.

To assess the performance of the proposed approach, we apply the trained network to reconstruct entanglement quasiprobabilities for two- and three-qubit systems from incomplete measurement data at varying sampling rates. Numerical simulations demonstrate that the reconstructed EQPs closely match the exact results obtained from the true density matrices, while requiring substantially fewer measurement settings than full quantum state tomography. This accuracy--cost advantage becomes increasingly pronounced with growing system size, highlighting the scalability of the method.

We benchmark the neural-network-based EQP reconstruction against two standard likelihood-based tomographic methods, regularized maximum likelihood estimation (MaxLik) and maximum-entropy maximum likelihood estimation (MLME) \cite{hradil1997quantum}. Both baselines infer the density matrix by optimizing a likelihood function and typically rely on complete measurement data. By contrast, our approach embeds measurement-basis information directly into the network input, enabling adaptive and accurate EQP reconstruction from incomplete measurements.

Across both two- and three-qubit systems, the neural network achieves near-unity reconstruction fidelity under sparse measurement conditions and consistently outperforms conventional tomographic methods as the system dimension increases. The reconstruction accuracy is quantified using the root-mean-square error (RMSE),
\begin{equation}
\mathrm{RMSE}=\sqrt{\frac{1}{n}\sum_{i=1}^{n}(y_i-x_i)^2},
\end{equation}
where $x_i$ and $y_i$ denote the exact and reconstructed EQP components, respectively.

\subsection{Two-Qubit System Validation}\label{subsection:2qubit}
We systematically benchmark the reconstruction accuracy of entanglement quasiprobabilities (EQPs) in a two-qubit system using three methods---EQPs-AIME-Net, MaxLik, and MLME---as a function of the number of measurement projectors (from 2 to 36 in steps of 2). Using the RMSE as the metric, we construct a strictly nested sequence of projector subsets $\{S_k\}_{k=1}^{18}$ with cardinalities $\lVert S_k\rVert=2k$, drawn from a universal set of 36 mutually unbiased bases via sequential sampling without replacement (Supplementary Sec.~S1). Starting from a uniform random draw, each subsequent subset is formed by sampling uniformly from the remaining pool, ensuring $S_k\subset S_{k+1}$. This nested design supports a rigorous multi-scale assessment of reconstruction robustness and directly addresses the minimal measurement budget required to reliably resolve nonclassical features in resource-constrained quantum tomography.

\begin{figure}[b]
\centering
\includegraphics[scale=1.0]{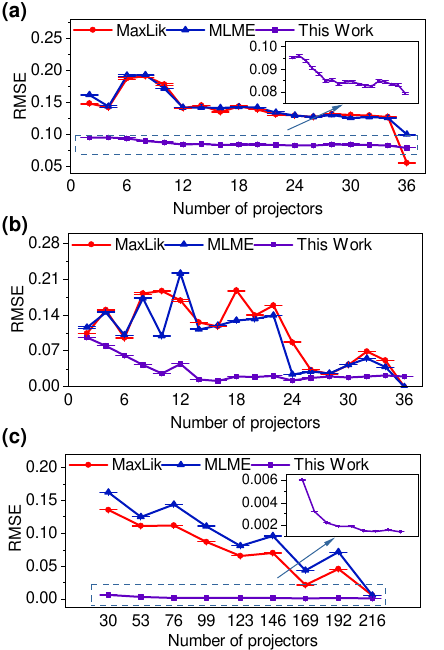}
\caption{\textbf{EQP reconstruction error versus measurement cost.} Root-mean-square error as a function of the number of projectors for (a) generic two-qubit states, (b) noisy Werner states, and (c) three-qubit systems. The deep-learning-based approach is shown by purple squares, standard maximum-likelihood estimation (MaxLik) by red circles, and maximum-entropy maximum-likelihood estimation (MLME) by blue triangles. Error bars indicate statistical uncertainty.}
\label{Result:EQPsLoss}
\end{figure}

As shown in Fig.~\ref{Result:EQPsLoss}(a), across the full measurement range, EQPs-AIME-Net achieves a $38$--$40\%$ lower mean RMSE than MaxLik and MLME ($0.0861$ vs.\ $0.1399$ and $0.1433$) and exhibits substantially improved stability (coefficient of variation $5.3\%$ vs.\ $20.9\%$ and $16.4\%$). Notably, EQPs-AIME-Net effectively leverages additional projectors without overfitting, reflected by its shallow RMSE trend (slope $-3.7\times10^{-4}$, $R^2=0.725$) compared with the steeper trends of MaxLik ($-1.98\times10^{-3}$) and MLME ($-1.78\times10^{-3}$). Crucially, in the low-projector regime ($\leq10$), EQPs-AIME-Net avoids the mid-range degradation observed in the baselines (peak RMSE $\approx0.19$ for 6--10 projectors), which arises from a transient constraint-inconsistency regime induced by heterogeneous projector measurements (Supplementary Sec.~S2). Although redundant measurements at larger projector counts partially suppress these artifacts and accelerate baseline convergence, their reconstructions remain less accurate and more variable than those of EQPs-AIME-Net.

At full projector counts, the minimal RMSE attained by MaxLik signals a saturation of EQPs-AIME-Net, with only marginal performance differences between methods. This indicates that in data-rich regimes, global statistical optimization can surpass learning-based approaches that are limited by training dynamics and model capacity. Error-propagation analysis shows that the remaining limitation of statistical methods is dominated by the intrinsic numerical sensitivity of EQPs, where small numerical perturbations are amplified.

To assess generalization, we evaluate the trained model on out-of-distribution Werner states, a family of two-qubit mixed states absent from the training set. Werner states are commonly used to represent experimental white noise and can be expressed as
\begin{equation}
   \rho_{\mathrm{W}}(p)=p\,\rho_{\psi^-}+\frac{1-p}{4}\mathbb{I},
\end{equation}
where $\rho_{\psi^-}=|\psi^-\rangle\langle\psi^-|$ is the maximally entangled Bell state and $\mathbb{I}$ is an identity matrix. The mixing parameter $p$ is sampled from a truncated normal distribution $\mathcal{N}(\mu,\sigma^2)$ restricted to $[0,1-\epsilon]$ to ensure strict positivity. Figure~\ref{Result:EQPsLoss}(b) shows the reconstruction RMSE of EQPs as a function of the number of measurement projectors.

For Werner states, EQPs-AIME-Net exhibits markedly enhanced robustness and accuracy compared with MLME and MaxLik. In the low-projector regime (2--10), where conventional estimators are highly sensitive to projector choice and yield large, fluctuating errors (RMSE $\sim0.1$--$0.2$), EQPs-AIME-Net rapidly converges from an initial RMSE of $0.0974$ to $0.0265$. In the intermediate regime (12--24), baseline methods improve substantially as redundancy increases, yet EQPs-AIME-Net already operates near its minimum error, indicating efficient extraction of entanglement features under shadow-like sampling. In the high-projector regime (26--36), MLME and MaxLik asymptotically approach zero RMSE, consistent with their optimality under complete data, while EQPs-AIME-Net saturates at $\sim0.02$. Notably, EQPs-AIME-Net maintains smooth error scaling ($\sigma=0.0246$ versus $\approx0.067$ for baselines) and bounded errors on out-of-distribution Werner states (maximum RMSE $<0.1$), avoiding the divergence ($>0.2$) observed for conventional methods. 

Overall, two-qubit results show that EQPs-AIME-Net achieves high accuracy and stability at low projector counts ($\leq10$). Its shallower regression slope at large projector numbers reflects a trade-off between strong generalization and optimal performance in data-rich regimes.
\begin{figure*}[htbp]
    \centering
    \includegraphics[width=0.75\textwidth]{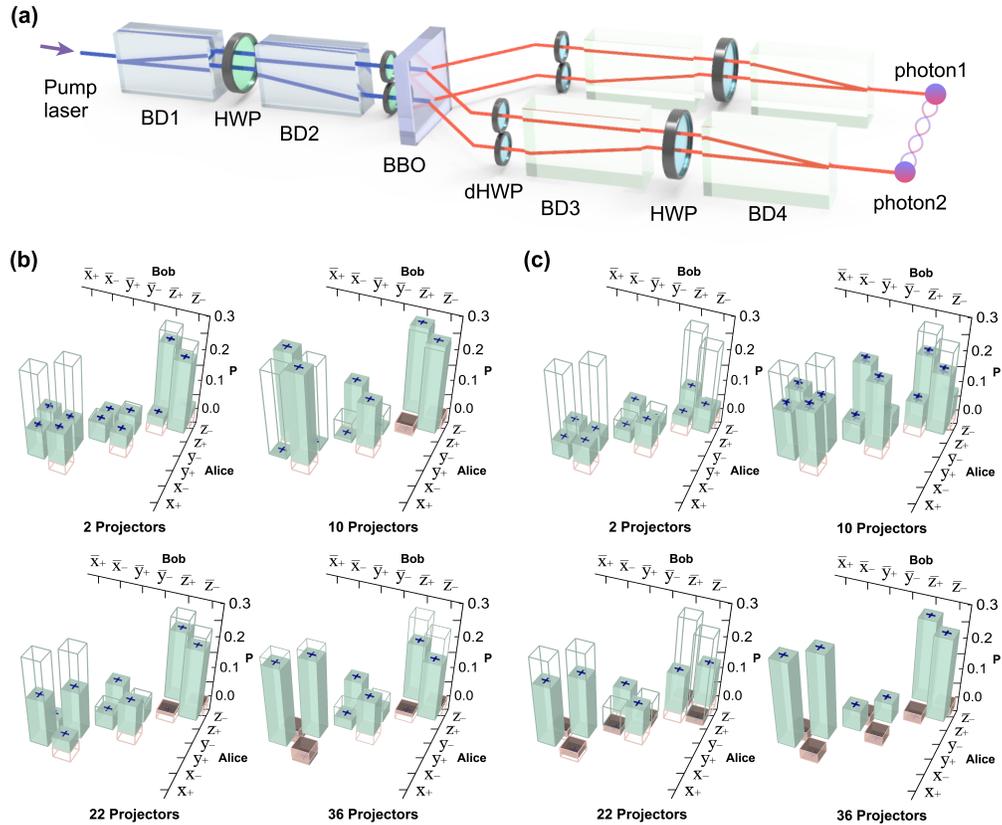}
    \caption{\textbf{Evolution of EQPs reconstruction under varying projector number in experimental data.} 
(a) A schematic diagram of our experimental setup for generation of entangled-photons. BD, beam displacer; HWP, half-wave plate; dHWP, dual-path half-wave plate; BBO, $\beta-$Barium Borate crystal. (b) The reconstructed EQP using our deep-learning method. (c) The reconstructed EQP using the maximum likelihood method at 2, 10, 22, and 36 projectors. The open columns and solid columns represent theoretical and experimental values, respectively.  Light coral red denotes negative components (indicating entanglement), and light olive green denotes positive components.}
    \label{fig:EQPs_prediction}
\end{figure*}

\subsection{Tripartite Qubit System Validation}
\label{result:3qubit}
In figure~\ref{Result:EQPsLoss}(c), we compares the reconstruction errors of three methods for predicting Pauli expectation values in a three-qubit system under progressively incomplete Pauli measurements, from full tomography (216 projectors) to highly sparse sampling (30 projectors). Although MLME and MaxLik attain near-optimal accuracy with complete data (RMSE $\approx 5.8$--$5.9\times10^{-3}$), their performance degrades sharply to RMSE $\approx 0.13$--$0.16$ in the sparse limit. In contrast, EQPs-AIME-Net maintains uniformly low errors across all regimes (RMSE $1.4$--$6.1\times10^{-3}$, mean $2.2\times10^{-3}$), corresponding to a $20$--$50\times$ improvement under severe measurement incompleteness.

This robustness arises from the end-to-end learning architecture, which circumvents the ill-conditioned global optimization over the 15-dimensional space of physical density matrices inherent to likelihood-based methods. By adaptively enhancing entanglement-sensitive features—especially high-weight correlators such as $\langle \sigma^{x}\!\otimes\!\sigma^{x}\!\otimes\!\sigma^{x} \rangle$—and exploiting quasi-probabilistic representations akin to discrete Wigner functions, EQPs-AIME-Net suppresses biases induced by reduced Pauli diversity. By contrast, MLME and MaxLik are susceptible to unstable likelihood landscapes and undersampling of many-body operators, leading to large variance and systematic infidelity.

It shows that EQPs-AIME-Net achieves substantially higher accuracy and robustness, with reconstruction errors as low as $\sim1/50$ of those of baseline methods. Scaling this approach to larger systems will require addressing the training and data-scaling challenges posed by the curse of dimensionality, motivating the development of few-shot or data-efficient learning strategies.

\subsection{\label{result:exp}Validation with a Nonlinear Photonic Entanglement Source}
To validate the proposed deep-learning model under realistic conditions, we apply it to experimental data from a two-qubit entangled-photon source. As shown in Fig.~\ref{fig:EQPs_prediction}(a), a pulsed laser at $775\,\mathrm{nm}$ is split into two parallel horizontally polarized beams using half-wave plates (HWPs) and beam displacers (BDs). The two beams are then focused onto the upper and lower regions of a $\beta-$Barium Borate crystal, where photon pairs at $1550\,\mathrm{nm}$ are generated via spontaneous parametric down-conversion (SPDC). The down-converted fields from the two regions are recombined by an inverse beam-displacing operation. By finely adjusting the BD tilt to achieve spatial and temporal compensation, the emission pathways become indistinguishable, enabling the preparation of a maximally entangled state. The output biphoton state is analyzed with polarization analyzers and detected by superconducting nanowire single-photon detectors. In the experiment, we prepare the Bell state $|\Phi^-\rangle$ with a measured fidelity of $0.99$ and record coincidence counts in multiple measurement bases.

Figure~\ref{fig:EQPs_prediction}(b) and (c) show the EQP distributions predicted by EQPs-AIME-Net and the maximum-likelihood method, respectively. With increasing projector number, both approaches gradually recover the correct sign structure that separates entangled negative components from separable positive ones. EQPs-AIME-Net converges smoothly from the low-projector regime with robust sign stability and largely avoids the spurious positive bias observed in the baseline. In contrast, the maximum-likelihood method exhibits strong fluctuations and frequent misclassification of negative components, requiring substantially more measurements for correction.

\subsection{\label{EQPsToOthers}From Reconstructed EQPs to Other Properties}

\begin{figure}[t]
\centering
\includegraphics[scale=1.1]{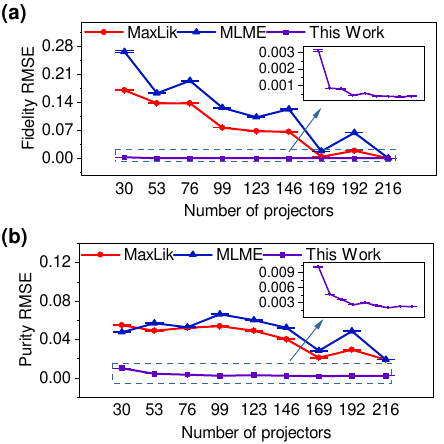}
\caption{\textbf{Performance comparison for predicting Fidelity (a) and Purity (b) from EQPs.} The RMSE of the deep learning approach trained on specific incomplete measurements is depicted by purple squares, the standard MaxLik method by red circles, and the MaxLik method with maximum entropy regularization by blue triangles.}
\label{fig:EQPsToOthers}
\end{figure}

In measurement-reduced entanglement characterization, quantities computed at high cost are often usable only once, since any reuse typically requires repeating measurements and recomputation. Multitask neural networks~\cite{Caruana1997Multitask} can target predefined observables, but encoder-based latent representations are usually hard to interpret and must be queried indirectly through learned mappings~\cite{Wu2024}. By contrast, EQPs provide an optimal and symbolic entanglement decomposition. Their coefficients reconstruct the density matrix $\rho$ via Eq.~(\ref{equ:1}), enabling inference of additional properties without storing the full measurement record. In this way, EQPs retain high-fidelity information about the state and its entanglement while alleviating the exponential storage burden, offering a compact and physically meaningful encoding for quantum information processing.

Here we compute EQPs using the incomplete-measurement protocol on the test set of tripartite qubit. From the reconstructed EQPs we obtain the density matrix $\rho$ and directly evaluate the fidelity and purity, which we benchmark against quantum state tomography,
\begin{align}
F(\rho,\sigma) &= \left(\mathrm{Tr}\sqrt{\sqrt{\rho}\,\sigma\,\sqrt{\rho}}\right)^2, \\
P(\rho) &= \mathrm{Tr}(\rho^2),
\end{align}
where $\sigma$ denotes the ideal reference state.

Figure~\ref{fig:EQPsToOthers} shows that our method preserves high accuracy as the number of projections is reduced. At 30 projections, the RMSE is $\sim 0.003$ for fidelity and $\sim 0.010$ for purity, substantially outperforming MaxLik and MLME, whose errors increase sharply in the sparse regime, with MaxLik reaching a fidelity RMSE of $0.170$. The numerical simulations for prediction of these quantum properties were implemented using the MindSpore Quantum framework~\cite{xu2024mindspore}, which seamlessly integrates quantum circuit simulation with deep learning-based optimization and automatic differentiation. These results indicate that EQPs-AIME-Net enables efficient density-matrix reconstruction from noisy and incomplete data, while providing reliable estimates of derived state properties without auxiliary post-processing. This capability is well suited to near-term tasks such as quantum error correction~\cite{Liao2024Machine,patil2024entanglemen} and quantum machine learning~\cite{Huang2025Direct}.

\section{Discussion}
We have proposed a deep-learning framework for incomplete and noise-robust prediction of entanglement quasiprobability distribution. This quasiprobability distribution provides an informationally complete representation and enables visualization and identification of entanglement. Compared with conventional tomographic methods, our approach achieves about a $40\%$ reduction in RMSE and a low coefficient of variation of $5.3\%$ in two-qubit systems. For three qubits, it suppresses the average reconstruction error by more than an order of magnitude and retains over a 20-fold advantage even with only $14\%$ of the full measurement set. The model generalizes reliably to diverse and out-of-distribution quantum states, and its practicality is confirmed by experiments with polarization-entangled photon pairs. The reconstructed quasiprobabilities can further be used to estimate derived state properties (Results, Fig.~\ref{fig:EQPsToOthers}), suggesting a unified route to entanglement assessment and property estimation from the same representation. These results suggest that EQPs-AIME-Net acts as an effective encoder of quantum systems, directly revealing entanglement through negative EQP components from highly limited data while simultaneously supporting faithful inference of other state properties.

The incomplete-measurement framework can be further combined with noise modeling to improve the characterization of realistic entangled states, and additional gains may be achieved by optimizing the weighting of sign-related terms in the loss function. A practical limitation is the computational cost of training EQPs-AIME-Net, which becomes substantial for large data sets and complex architectures and may restrict near-term deployment in resource-limited settings. Progress in AI hardware and the development of efficient, universal deep-learning frameworks~\cite{xu2024mindspore} are expected to mitigate these constraints.

More broadly, this work pave a way toward resource-efficient entanglement characterization in the noisy intermediate-scale quantum regime.  Its ability to perform reliable EQP prediction under noisy and incomplete measurements enables real-time applications in quantum computing \cite{raussendorf2001one} and quantum networks \cite{azuma2023quantum} where entanglement is a key operational resource. The experimental validation under realistic noise conditions further indicates the readiness of this method for laboratory use, providing a practical tool for certifying and exploiting quantum resources.

\section{\label{sec:methods}Materials and methods}
\subsection{Entanglement Quasiprobabilities and Nonclassicality}
Quantum states are described by density operators $\hat{\rho}$, which are commonly expressed as convex mixtures of pure states. Here, we introduce a specialized classical decomposition of the form
\begin{equation}
\hat{\rho} = \int dP_{\mathrm{cl}}(c)\, |c\rangle\langle c|,
\label{equ:1}
\end{equation}
where $|c\rangle$ ($c \in \mathcal{C}$) denote pure states associated with configurations of a prescribed classical model $\mathcal{C}$, and $P_{\mathrm{cl}}(c)$ is a corresponding quasiprobability distribution. A state is deemed classical if $P_{\mathrm{cl}}(c)\ge 0$ for all $c$, and nonclassical otherwise. This construction generalizes the quasiprobability formalism of quantum optics, where negativities constitute a definitive signature of nonclassical light~\cite{weedbrook2012gaussian}.  

Equivalently, a quantum state can be decomposed as
\begin{equation}
\hat{\rho}
=\sum_{n} p_{n} \lvert c_{n} \rangle \langle c_{n} \rvert
+\hat{\rho}_{\mathrm{res}},
\end{equation}
where $\lvert c_{n} \rangle\langle c_{n} \rvert \in \mathcal{C}$. The state is classical if $p_{n}\ge 0$ for all $n$ and $\hat{\rho}_{\mathrm{res}}=0$, while nonclassicality is revealed by negative coefficients or a nonvanishing residual term. The entanglement quasiprobability framework seeks an optimal decomposition within $\mathrm{conv}(\mathcal{C})$, which is achieved by identifying a finite set of extremal stationary points $\mathcal{D}\subset\mathcal{C}$ associated with $\hat{\rho}$~\cite{Sperling2018Quasiprobability}.

Any $\hat{\rho}\in \mathrm{conv}(\mathcal{C})$ admits a convex decomposition in terms of the stationary points $\mathcal{D}$, leading to a linear system for the quasiprobabilities $\vec{p}$,
\begin{equation}
G\,\vec{p}=\vec{g},
\end{equation}
where $\vec{g}$ collects the scalar products $(\hat{\rho}|\hat{d})$ with $\hat{d}\in\mathcal{D}$, $\vec{p}$ contains the corresponding weights, and $G$ is the Gram matrix formed by inner products between elements of $\mathcal{D}$. For classical states, at least one solution with $\vec{p}\ge 0$ exists.

The construction proceeds as follows. First, elements of the classical set $\mathcal{C}$ are parametrized as a differentiable map $t\mapsto\hat{\Gamma}(t)\in\mathcal{C}$, where $t$ may be multidimensional. Stationary points are then obtained by solving $\partial_{t}(\hat{\rho}|\hat{\Gamma}(t))=0$, together with possible isolated solutions, yielding the set $\mathcal{D}$. Solving the linear system $G\vec{p}=\vec{g}$ provides the quasiprobabilities. A state is identified as nonclassical if no solution satisfies $p_{d}\ge 0$ for all $d\in\mathcal{D}$, or if a nonvanishing residual $ \hat{\rho}_{\mathrm{res}} =\hat{\rho}-\sum_{d\in\mathcal{D}} p_{d}\,\hat{d}$ remains.

Conversely, once the entanglement quasiprobabilities are determined, the density operator $\hat{\rho}$ can be reconstructed according to Eq.~(\ref{equ:1}), enabling the direct inference of additional global and local properties of the quantum system.

\subsection{\label{sec:DNN}Deep Learning Methodology}
EQPs-AIME-Net follows the paradigm of measurement-specific incomplete-measurement networks, where the input consists of both the observed outcome probabilities and encoded representations of the corresponding measurement projectors. This dual-input design enables the direct inference of entanglement properties under partial measurement constraints, while retaining access to both local and global information of the quantum system for two-qubit and multipartite states.

To address the high-dimensional input spaces associated with entangled states, the network incorporates residual connections that stabilize training and mitigate gradient degradation in deep architectures. This design alleviates the curse of dimensionality encountered with sparse and noisy measurement data, allowing subtle entanglement signatures to be faithfully captured. As a result, EQPs-AIME-Net reconstructs full-system entanglement indicators from incomplete observations and generalizes effectively to larger Hilbert spaces, as demonstrated in our two- and three-qubit experiments.

In preliminary two-qubit experiments, a standard multilayer perceptron (MLP) achieves low EQP prediction errors by capturing nonlinear correlations in moderately sized inputs. However, extending this architecture to three-qubit systems—where both the number of measurement projectors and their parametrization grow exponentially—leads to severe training instabilities, including gradient degradation and poor generalization. These failures originate from noise amplification and informational sparsity inherent to incomplete measurements in rapidly expanding Hilbert spaces, rendering conventional feed-forward networks ineffective for resolving multipartite entanglement.

In EQPs-AIME-Net, shortcut connections enable stable gradient propagation, while controlled parameter scaling ensures robust convergence and enhanced reconstruction fidelity. This design enables scalable entanglement reconstruction from incomplete measurements, addressing a central challenge in quantum machine learning and facilitating faithful quantum-state representation beyond few-qubit regimes.
\subsection{Training Data and Protocol}
To train and evaluate EQPs-AIME-Net at the two- and three-qubit levels, scale-specific datasets were constructed. For the two-qubit case, $3.05\times10^{5}$ quantum states were generated, with $80\%$ sampled from the Bures ensemble to ensure broad coverage of the mixed-state space. These states are given by
\begin{equation}
\rho = \frac{(1 + U^\dagger) G G^\dagger (1 + U)}{\mathrm{Tr}[(1 + U^\dagger) G G^\dagger (1 + U)]},
\end{equation}
where $G$ is a complex Ginibre matrix with entries $G_{ij}\sim \mathcal{N}(0,1)+i\mathcal{N}(0,1)$ and $U$ is drawn from the Haar measure. The remaining $20\%$ consist of Haar-random pure states admixed with white noise. An independent test set of $5\times10^{3}$ states was reserved for final evaluation, while the remaining data were split into training and validation sets in a $4{:}1$ ratio.

To assess generalization beyond the training distribution, the trained two-qubit model was further tested on states drawn from an out-of-distribution ensemble, with generation details and results reported in Sec.III.

For multipartite systems, we considered a parameterized family of $N$-qubit states of the form
\begin{equation}
\hat{\rho} = \frac{1}{2^N}\!\left(\hat{\sigma}_0^{\otimes N}
+ \rho_z \hat{\sigma}_z^{\otimes N}
+ \rho_x \hat{\sigma}_x^{\otimes N}
+ \rho_y \hat{\sigma}_y^{\otimes N}\right),
\end{equation}
where all physically valid combinations of $(\rho_x,\rho_y,\rho_z)\in[-1,1]^3$ were explored via grid discretization. From the resulting ensemble, $9.4\times10^{4}$ distinct quantum states were selected. An independent test set of $4\times10^{3}$ states was held out, while the remaining $9.0\times10^{4}$ states were split into training and validation sets in a $4{:}1$ ratio.
\subsection{Nonlinear optical setup for generation of entangled photons}
We generate polarization-entangled photon pairs via an ultrafast nonlinear spontaneous parametric down-conversion process. A 775 nm femtosecond pump laser (pulse width $\approx 150$ fs, repetition rate of 80 MHz) is first adjusted to $+45^{\circ}$ linear polarization using a half-wave plate, is evenly split into two beams by a beam displacer. After polarization flipping, the separation between the two beams is increased to 2.6 mm. Both beams are then incident on a same 6 mm thick BBO crystal for a Type-II SPDC process, aligned for producing wavelength-degenerate photon pairs $|H\rangle_1|V\rangle_2$ at 1550 nm. 
Following independent polarization adjustment using dual-path half-wave plates, the photon pairs are directed through a reversed-path superposition setup.
When only one photon pair is generated for one pump pulse, polarization entanglement is achieved due to the fundamental indistinguishability of the photon pair’s origin (upper path for generation of $|H\rangle_1|H\rangle_2$ or lower path for $|V\rangle_1|V\rangle_2$). The polarization state of each photon is analyzed using a quarter-wave plate, a half-wave plate, and a polarizing beam splitter, enabling arbitrary projective measurements. Both photons are collected by lenses with a focal-length of 300 mm, with residual pump light filtered out using a silicon filter. Detection is performed using superconducting nanowire single-photon detectors with a system efficiency of approximately 85\%.   Coincidence counting and analysis reveal a coincidence efficiency exceeding 50\%.

\subsection*{Acknowledgments}
This work is sponsored by CPS-Yangtze Delta Region Industrial Innovation Center of Quantum and Information Technology-MindSpore Quantum Open Fund. We are grateful for the financial supports from the National Natural Science Foundation of China (Grant No. 62405025), the China Postdoctoral Science Foundation (Grant No. 2024M754109) and the Zhejiang Province Postdoctoral Research Funding (ZJ2024096). We also thank Zidu Liu for valuable suggestions of this work. Y.-Z. L. and L.-C. P. contributed equally to this work. 
\subsection*{Data availability}
The code used to generate the results of this study is available at \url{https://github.com/PhotonicsQuantum-Alex/EQPs-AIME-Net}. The data that supports the plots within this paper and other findings of this study are available from the corresponding author upon reasonable request.
\section*{Conflict of interest}
The authors declare no competing interests.

\bibliography{apssamp}

\end{document}